\pdfoutput=1
\documentclass[aps,showpacs,twocolumn,twoside,prl,superscriptaddress]{revtex4-1}
\usepackage{epsfig,amsmath,amssymb}
\usepackage{graphicx}
\usepackage{array}
\usepackage{xcolor}

\def\={\!=\!}
\def\>{\!>\!}
\def\<{\!<\!}
\def\-{\!-\!}
\def\+{\!+\!}

\def\uvo#1{\lq\lq #1\rq\rq}

\begin{document}

\title{Complex density of continuum states in resonant quantum tunneling}

\author{Pavel Str\'ansk\'y}
\email{stransky@ipnp.mff.cuni.cz}
\affiliation{Institute of Particle and Nuclear Physics, Faculty of Mathematics and Physics, Charles University, V Hole\v{s}ovi\v{c}k\'ach 2, 18000 Prague, Czech Rep.} 
\author{Milan {\v S}indelka}
\affiliation{Institute of Plasma Physics, Academy of Sciences of the Czech Republic, Za Slovankou 3, 18200 Prague, Czech Rep.} 
\author{Michal Kloc}
\affiliation{Institute of Particle and Nuclear Physics, Faculty of Mathematics and Physics, Charles University, V Hole\v{s}ovi\v{c}k\'ach 2, 18000 Prague, Czech Rep.}
\affiliation{Department of Physics, University of Basel, Klingelbergstrasse 82, 4056 Basel, Switzerland} 
\author{Pavel Cejnar}
\email{cejnar@ipnp.mff.cuni.cz}
\affiliation{Institute of Particle and Nuclear Physics, Faculty of Mathematics and Physics, Charles University, V Hole\v{s}ovi\v{c}k\'ach 2, 18000 Prague, Czech Rep.}

\date{\today}

\begin{abstract}
We introduce a complex-extended continuum level density and apply it to one-dimensional scattering problems involving tunneling through finite-range potentials.
We show that the real part of the density is proportional to a real \uvo{time shift} of the transmitted particle, while the imaginary part reflects the imaginary time of an instanton-like tunneling trajectory. 
We confirm these assumptions for several potentials using the complex scaling method. 
In particular, we show that stationary points of the potentials give rise to specific singularities of both real and imaginary densities which represent close analogues of excited-state quantum phase transitions in bound systems. 
\end{abstract}

\maketitle

The density of discrete energy spectra of bound quantum systems forms a bridge between classical and quantum mechanics
\cite{Gut71,Bal74,Ber76}.
While the oscillatory component of the level density is related to classical periodic orbits, the smooth component is determined by the size of the accessible phase space at given energy~$E$.
Non-analyticities of the phase-space volume function lead to singularities of quantal spectra known as the excited-state quantum phase transitions (ESQPTs), see e.g. Refs.\,\cite{Cej06,Cap08,Lar13,Str14+,Bas14a,Bas14,Rel14,Kop15,Str16,Gra16,Sin17,Die17,Byc18,Gar18,Hum19,Kha19,Mac19}.
In particular, for one-dimensional (1D) systems, these singularities express anomalous time relations caused by discontinuous or divergent periods of classical orbits.

The question addressed in this work is whether the densities of continuum energy eigenstates in {\em unbound \/} systems allow for similar semiclassical interpretations and show analogous ESQPT effects.
This concerns fundamental scattering and tunneling processes that include molecules, atoms, nuclei and elementary particles. 
Time relations in such processes and their semiclassical foundations are vividly discussed topics \cite{Wig55,Smi60,Tsa75,Lan94,Car02,Sok18}, which with recent advent of the attosecond metrology become available to experimental study, see e.g. Refs.\,\cite{Hen01,Sha12,Lan15,Sat19}.

We focus on 1D scattering problems involving resonant tunneling in multibarrier potentials \cite{Raz14}.
Quantum tunneling, in general, with its possible role in the evolution of early universe \cite{Col77a} and indisputable impact on modern technologies \cite{Nobel}, is thought to be a genuinely quantum phenomenon with no classical counterpart.
However, its analogues have been searched in terms of a {\em complex generalization\/} of classical mechanics \cite{Lau72,Col77,Mai97,Tak99,Deu10+}.
Our present work brings new arguments supporting these efforts.
We find that a suitable complexly defined continuum level density reflects complex times deduced from generalized semiclassical dynamics involving instanton-like solutions.  
Moreover, we demonstrate that real and imaginary components of the continuum level density show ESQPT-like singularities associated with classical stationary points of potentials $V(x)$ and $-V(x)$ applied, respectively, in the allowed and forbidden regions.

Let $\hat{H}$ and $\hat{H}^{(0)}$ be Hamiltonians with continuous energy spectra describing an unbound quantum system with and without interaction, respectively.
The continuum level density used in the corresponding scattering problems has been defined \cite{Lev69,Kru98+} as 
\begin{equation}
\delta\rho(E)\=-\frac{1}{\pi}\lim_{\epsilon\to 0}{\rm Im}{\rm Tr}\left[\hat{G}(E\+i\epsilon)-\hat{G}^{(0)}(E\+i\epsilon)\right]
\label{olde},
\end{equation} 
where $\hat{G}(E)\=1/(E\-\hat{H})$ and $\hat{G}^{(0)}\=1/(E\-\hat{H}^{(0)})$ are Green operators associated with $\hat{H}$ and $\hat{H}^{(0)}$. 
Our present approach differs from this in two points: (a) we assume that both Hamiltonians have discrete sets of complex eigenvalues and (b) we introduce the continuum level density $\Delta\rho({\cal E})$ as a complex function in the complex energy domain ${\cal E}\=E\-\frac{i}{2}\Gamma$.
Point (a) is realized by application of the so-called complex scaling method in combination with a finite-box approximation \cite{Bal71,Sim72+,Ho83,Moi98,Suz05+,Moi11}.
The method makes use of a similarity transformation with a non-unitary operator $\hat{S}$, which maps the original Hamiltonians to equivalent non-Hermitian images $\hat{H}_{\rm NH}\=\hat{S}\hat{H}\hat{S}^{-1}$ and $\hat{H}^{(0)}_{\rm NH}\=\hat{S}\hat{H}^{(0)}\hat{S}^{-1}$.
For a finite box size, these images have discrete sets of complex eigenvalues ${\cal E}_k=E_k\-\frac{i}{2}\Gamma_k$ and ${\cal E}^{(0)}_k=E^{(0)}_k\-\frac{i}{2}\Gamma^{(0)}_k$ with integer index $k$. 
As discussed below, some of these states can be interpreted as resonance states with real energies $E_k$ and widths $\Gamma_k\!\geq\!0$.
Generalization (b) is achieved by defining the continuum level density as
\begin{equation}
\Delta\rho({\cal E})=\underbrace{\frac{i}{\pi}\,{\rm Tr}\,\frac{1}{{\cal E}-\hat{H}_{\rm NH}}}_{\rho({\cal E})}-\underbrace{\frac{i}{\pi}\,{\rm Tr}\,\frac{1}{{\cal E}-\hat{H}^{(0)}_{\rm NH}}}_{\rho^{(0)}({\cal E})}
\label{lev},
\end{equation}
where the traces are evaluated as sums over all discrete eigenvectors, whose energies  ${\cal E}_k$ and ${\cal E}^{(0)}_k$ represent poles of $\Delta\rho({\cal E})$. 
In the infinite-box limit, the real part of Eq.\,\eqref{lev} on the real energy axis, ${\rm Re}\Delta\rho(E\-i0)\!\equiv\!{\rm Re}\Delta\rho(E)\={\rm Re}\rho(E)\-{\rm Re}\rho^{(0)}(E)$, has to coincide with density \eqref{olde}. 
The meaning of ${\rm Im}\Delta\rho(E)$ is discussed below.

Let us briefly overview properties of the density \eqref{lev}.
From the residue theorem we see that a contour integral of $\Delta\rho({\cal E})$ along a closed loop in the complex plane ${\cal E}$ gives twice the difference between the number of eigenvalues of $\hat{H}_{\rm NH}$ and $\hat{H}_{\rm NH}^{(0)}$ inside the loop.
We expect that ${\cal E}_k\!\approx\!{\cal E}^{(0)}_k$ for $E$ much larger than the energy range of interaction $\hat{V}\=\hat{H}\-\hat{H}^{(0)}$, so contributions of these eigenstates to Eq.\,\eqref{lev} approximately cancel out.
The essential part of $\Delta\rho({\cal E})$ therefore comes from a finite number (in the finite-box approximation) of eigenstates at smaller energies $E$. 
The $\rho({\cal E})$ term in Eq.\,\eqref{lev} reads
\begin{eqnarray}
{\rm Re}\rho({\cal E})&=&\frac{1}{\pi}\sum_k\frac{-\frac{1}{2}(\Gamma\-\Gamma_k)}{(E\-E_k)^2+\frac{1}{4}(\Gamma\-\Gamma_k)^2},
%-\frac{1}{\pi}\sum_k\frac{\frac{1}{2}(\Gamma^{(0)}_k\-\Gamma)}{(E\-E^{(0)}_k)^2+\frac{1}{4}(\Gamma\-\Gamma^{(0)}_k)^2}
\label{rer}\\
{\rm Im}\rho({\cal E})&=&\frac{1}{\pi}\sum_k\frac{E\-E_k}{(E\-E_k)^2+\frac{1}{4}(\Gamma\-\Gamma_k)^2},
\label{imr}
\end{eqnarray}
and the $\rho^{(0)}({\cal E})$  term is expressed analogously. 
Below we will analyze $\Delta\rho({\cal E})$ on the real energy axis, i.e., for $\Gamma\=0$. 
The real part of $\rho(E)$ represents a generalization of the level density $\varrho(E)=\sum_k\delta(E\-E_k)$ of a bound system with discrete energies $E_k$ to the smooth form ${\rm Re}\rho(E)\=\sum_k\delta_{\Gamma_k}(E\-E_k)$, where $\delta_{\Gamma}(\Delta E)\=\frac{1}{2\pi}\Gamma/(\Delta E^2\+\frac{1}{4}\Gamma^2)$ is a normalized Breit-Wigner peak (Cauchy distribution) with the maximum at $\Delta E\=0$ and the full width at half-maximum $\Gamma$.
An analogous expression applies to ${\rm Re}\rho^{(0)}(E)$, so ${\rm Re}\Delta\rho(E)$ consists of positive and negative peaks centered at energies $E_k$ and $E_k^{(0)}$, respectively.
If the widths $\Gamma_k$ are close to zero, an additional smoothening may be needed to get rid  of sharp local structures and reveal a robust energy dependence of the level density.
This can be achieved by adding a small positive imaginary component $i\epsilon$ to energy $E$, i.e., by setting $\Gamma\=-2\epsilon$ in Eqs.\,\eqref{rer}, \eqref{imr} and their $\rho^{(0)}(E)$ analogues.
Hence we introduce smoothed level densities $\rho(E\+i\epsilon)\!\equiv\!{\overline\rho}(E)$, $\rho^{(0)}(E\+i\epsilon)\!\equiv\!{\overline\rho}^{(0)}(E)$ and  
$\Delta{\overline\rho}(E)\={\overline\rho}(E)-{\overline\rho}^{(0)}(E)$.

The complex level  density \eqref{lev} will be investigated in 1D scattering problems. 
Hamiltonians of these problems have the standard forms, $\hat{H}^{(0)}\=\hat{p}^2/2m$ and $\hat{H}\=\hat{H}^{(0)}\+\hat{V}(x)$, where $\hat{p}\=-i\hbar\frac{\partial}{\partial x}$ stands for the momentum operator, $m$ for the particle mass, and $\hat{V}(x)$ is a potential.
We assume that $\hat{V}(x)\!\approx\!0$ outside a certain finite interval $(x_{\rm L},x_{\rm R})$. 
The usual asymptotics of wave functions is required, namely 
\begin{equation}
\psi(x)=\left\{\begin{array}{ll}
e^{+\frac{i}{\hbar}px}\+R(E)e^{-\frac{i}{\hbar}px} & {\rm for\ }x\<x_{\rm L},
\\
T(E)e^{+\frac{i}{\hbar}px} & {\rm for\ }x\>x_{\rm R},
\end{array}\right.
\label{asy}
\end{equation} 
where $p\=\sqrt{2mE}$, and $R(E)$ and $T(E)$ stand for reflection and transition amplitudes, respectively.
The transmission amplitude is written as
\begin{equation}
T(E)\=|T(E)|e^{i\phi(E)}\=e^{ i\left[\phi(E)-i\ln|T(E)|\right]}\equiv e^{i\Phi(E)}
\label{tra},
\end{equation}
where $\phi(E)$ is a real phase shift of the transmitted wave and $\Phi(E)$ is a complex phase. 

It is known that the real continuum level density \eqref{olde} expresses the change of the real phase shift $\phi(E)$ with energy \cite{Lev69}. 
In analogy, we write 
\begin{equation}
\Delta\rho(E)=\frac{1}{\pi}\frac{d}{dE}\Phi(E)=\frac{1}{\pi}\frac{d}{dE}\phi(E)\-\frac{i}{\pi}\frac{d}{dE}\ln|T(E)|.
%\quad\ 
\label{deriphas}
\end{equation}
This implies that the \uvo{observable} functions $\phi(E)$ and $|T(E)|$ can be obtained by integration of ${\rm Re}\Delta\rho(E)$ and ${\rm Im}\Delta\rho(E)$, respectively.
So the real and imaginary parts of the continuum level density jointly contain complete information on the transmission amplitude, and both are in principle accessible to experimental study.
Using $\Delta{\overline\rho}(E)\!\equiv\!\Delta\rho(E\+i\epsilon)$ instead of $\Delta\rho(E)$, one obtains a smoothed complex phase ${\overline\Phi}(E)\equiv{\overline\phi}(E)\-i\ln|{\overline T}(E)|$.

Phase shifts of wave functions in scattering problems are related to some suitably defined \uvo{time delays} \cite{Wig55,Smi60,Tsa75,Lan94,Car02,Sok18}.
For example, the so-called Eisenbud-Wigner time \cite{Wig55} is given as $\Delta t\=\hbar\frac{d}{dE}\phi$, which near the center of a single resonance yields a delay $\Delta t\!\propto\!\hbar/\Gamma_k$ proportional to the average lifetime, whereas far from the resonances $\Delta t\!\approx\!0$ \cite{Car02}.
To avoid sharp changes of the time delay, we calculate it from the smoothed phases. 
The complex phase ${\overline\Phi}(E)$ results in the complex time shift 
\begin{equation}
\Delta{\cal T}(E)\equiv\hbar\frac{d}{dE}{\overline\Phi}(E)
\label{cot}.
\end{equation}
The meaning of real and imaginary components of $\Delta{\cal T}(E)$ follows from semiclassical considerations.
Indeed, the transmitted wave at $x\=x_{\rm R}$ is approximated by 
\begin{equation}
{\overline T}(E)e^{\frac{i}{\hbar}px_{\rm R}}= e^{\frac{i}{\hbar}px_{\rm L}}\ e^{\tfrac{i}{\hbar}\left[\int\limits_{x_{\rm L}}^{x_{\rm R}}dx' \sqrt{2m[E-V(x')]}+C\right]},
\label{wkb}
\end{equation}
where constant $C$ includes phase shifts between allowed and forbidden regions \cite{Ber72}.
From Eq.\,\eqref{cot} we obtain
\begin{eqnarray}
{\rm Re}\Delta{\cal T}(E)&=&\!\!\!\!\!\!\!\!\int\limits_{E\geq V(x)}\!\!\!\!\!\!\!dx' \sqrt{\frac{m}{2[E\-V(x')]}}-\sqrt{\frac{m}{2E}}(x_{\rm R}\-x_{\rm L}),
\quad\label{ret}\\
{\rm Im}\Delta{\cal T}(E)&=&\!\!\!\!\!\!\!\!\int\limits_{E<V(x)}\!\!\!\!\!\!\!dx' \sqrt{\frac{m}{2[V(x')\-E]}}
\label{imt}.
\end{eqnarray}
The integral in Eq.\,\eqref{ret} is taken across all classically allowed (for given $E$) regions between $x_{\rm L}$ and $x_{\rm R}$ and represents the time that a classical particle needs to pass these regions.
The subtracted term is the transmission time of a free particle.
On the other hand, the integral in Eq.\,\eqref{imt}, taken across all classically forbidden regions, is related to complex-time solutions of the classical equations of motions inside the potential barriers.

The use of complex time in the description of tunneling processes within the framework of the path integral was initiated by McLaughlin \cite{Lau72} and later developed e.g. in Refs.\,\cite{Bal74,Col77,Mai97,Tak99,Deu10+}. 
In particular, the well known instanton solution \cite{Col77} applies the Wick rotation $t\!\to\!-it$ to derive the semiclassical tunneling probability.
This approach was generalized to multibarrier tunneling, for which a notrivial evolution of time ${\cal T}$ in the complex plane was considered \cite{Deu10+}.
In these problems, the continuous path ${\cal T}(s)$, characterized by a linearly increasing real parameter $s$, has a shape of a descending staircase, whose segments corresponding to motion in classically allowed regions are parallel with the real time axis ($\frac{d}{ds}{\cal T}\=1$), while the segments associated with tunneling through forbidden regions go vertically along the negative imaginary axis ($\frac{d}{ds}{\cal T}\=-i$).
Complex-extended Hamilton equations render the momentum switching between pure real and imaginary values at classical turning points between allowed and forbidden regions, while the coordinate evolves solely in the real domain.
The evolution in forbidden regions is equivalent to the motion with energy $-E$ in an inverted potential $-V(x)$.

These considerations lead to a semiclassical expression of the continuum level density in 1D scattering.
Combining Eqs.\,\eqref{deriphas} and \eqref{cot}, we get the formula 
\begin{equation}
\Delta{\overline\rho}(E)=\frac{1}{\pi\hbar}\Delta{\cal T}(E)
\label{deti},
\end{equation}
which together with Eqs.\,\eqref{ret} and \eqref{imt} represents a semiclassical estimate of both real and imaginary parts of the smoothed density.
The real and imaginary parts of the time shift $\Delta{\cal T}(E)$ correspond to real and imaginary times accumulated in the above-described staircase evolution of ${\cal T}(s)$, the real part being reduced by subtracting the passage time of a free particle.
We note an apparent similarity of Eq.\,\eqref{deti} with the relation ${\overline\varrho}(E)=t_{\circ}(E)/2\pi\hbar$ between the smoothed level density of a bound system and the period $t_{\circ}$ of classical orbits at energy $E$. 
The denominators in these formulas differ by factor 2 because the time shift $\Delta{\cal T}(E)$ includes only a half of the return trajectory.
Therefore, the scattering and bound systems rely upon rather similar semiclassical descriptions, in which the scattering systems with continuous energy require the use of complex instead of real quantities. 

\begin{figure}[t]
\includegraphics[width=\linewidth]{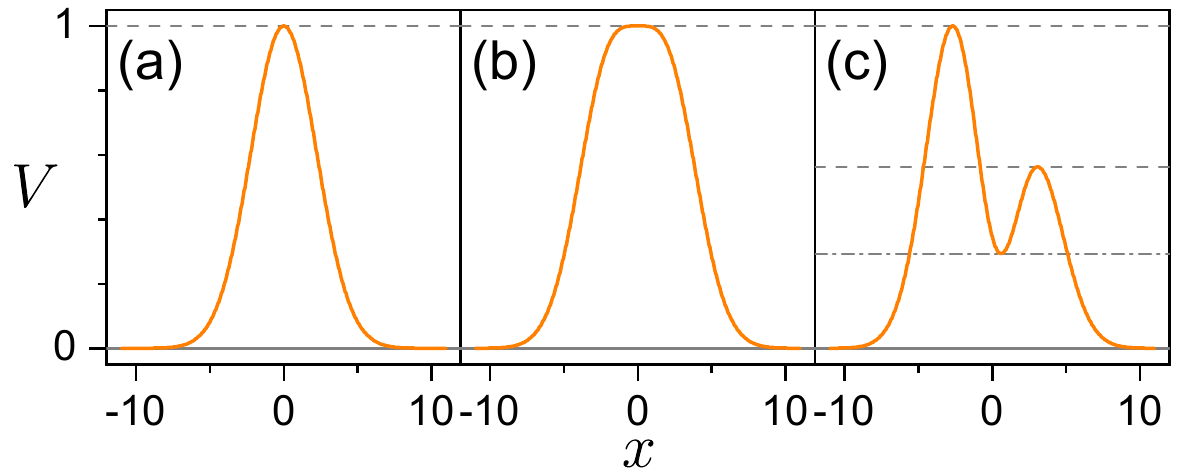}
\caption{
Potentials of the form \eqref{pot} with $(a, b, c, \eta)$ set to (a) $(1,0,0,\frac{1}{10})$, (b) $(1,0,\frac{1}{10},\frac{1}{10})$ and (c) $(0.346,-0.173,0.173,\frac{1}{10})$.
Dashed lines mark quadratic [in panels (a) and (c)] and quartic [in panel (b)] stationary points.}
\label{poti}
\end{figure}   

\begin{figure}[t]
\includegraphics[width=\linewidth]{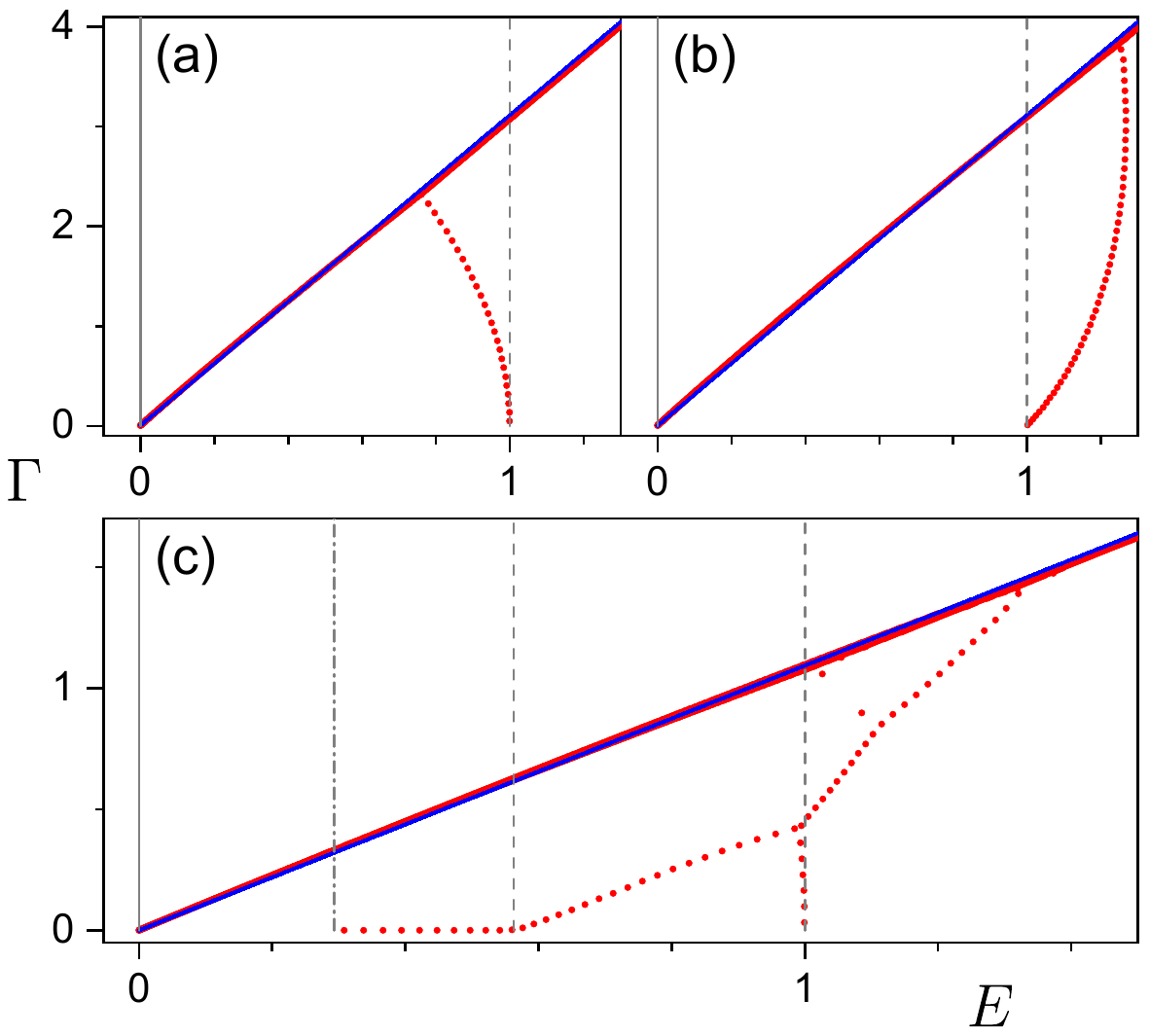}
\caption{
Resonances and background states calculated by the complex scaling method for potentials from the respective panels of Fig.\,\ref{poti}.
Red dots are complex eigenvalues of $\hat{H}_{\rm NH}$, blue dots those of $\hat{H}_{\rm NH}^{(0)}$.
Vertical lines mark energies of stationary points.
Parameters $(\hbar/\sqrt{m},\theta)$ are $(0.1,0.5)$ in panels (a) and (b), and $(0.058,0.25)$ in  (c).
The box size is $L=500$, for diagonalization we use $1.5\cdot10^4$ basis box functions.}
\label{reso}
\end{figure}   

We test Eq.\,\eqref{deti} in 1D scattering systems with several sample potentials (see Fig.\,\ref{poti}) using the complex scaling method \cite{Bal71,Sim72+,Ho83,Moi98,Suz05+,Moi11}. 
It is based on a similarity transformation $\hat{S}\=e^{i\theta/2}e^{-\theta\hat{x}\hat{p}/\hbar}$ with parameter $\theta\!\in\!(0,\frac{\pi}{4})$ which maps the coordinate and momentum operators to $e^{i\theta}\hat{x}$ and $e^{-i\theta}\hat{p}$.
This turns resonant solutions $\psi_k(x)$ associated with poles of the scattering matrix at complex momenta $p_k\=|p_k|e^{-i\alpha_k}$ with $\alpha_k\!\in\!(0,\theta)$ into square-integrable eigenstates of the transformed Hamiltonian $\hat{H}_{\rm NH}$.
The results of the complex scaling method for Hamiltonians with potentials from Fig.\,\ref{poti} are shown in Fig.\,\ref{reso}.
For a fixed $\theta$, only the resonances with complex energies $E_k\=|E_k|e^{-i\beta_k}$ satisfying $\beta_k\!\in\!(0,2\theta)$ emerge (red dots below the diagonal), while those with $\beta_k\!\geq\!2\theta$ form background states (in the infinite-size limit forming so-called \uvo{rotated continuum}) along the line $\frac{1}{2}\Gamma/E\=\tan 2\theta$ (red dots along the diagonal).
Resonances with larger $\beta_k$ can be found if $\theta$ is further increased, but this is numerically demanding and not essential here.
The effect of undisclosed resonances on the level density is included in the contribution of the background states for a given $\theta$.
Moreover, the complex scaling transformation performed on $\hat{H}^{(0)}$ yields only the background states (blue dots along the diagonal), and for large enough $E$ the background contributions in ${\rm Re}\rho(E)$ and ${\rm Re}\rho^{(0)}(E)$ approximately cancel each other \cite{Suz05+}. 
So ${\rm Re}\Delta\rho(E)$ is mostly formed by resonances not too far from the real axis, and the same holds for ${\rm Im}\Delta\rho(E)$.
Close to $E\=0$, both background contributions combine in a nontrivial way. 

The potentials employed here have a general form
\begin{equation}
V(x)=\left(a+bx+cx^2\right)e^{-\eta x^2},
\label{pot}
\end{equation}
where $a,b,c$ and $\eta$ are adjustable parameters.
We choose only three testing examples shown in Fig.\,\ref{poti}.
Calculations are performed with some optimized values of the classicality parameter $\hbar/\sqrt{m}$ and angle $\theta$ (see the caption of Fig.\,\ref{reso}).
For the sake of simplicity all quantities are taken dimensionless. 
In the finite-box approximation we assume that $V(x)\=\infty$ for $|x|\>\frac{1}{2}L$. 
The finite length $L$ makes the set of background states discrete, but its value is chosen large enough to keep this set dense and to yield the resonances at $\beta_k\<2\theta$ stabilized (invariant under an increase of $L$).
Though the potentials \eqref{pot} are not restricted to any finite support interval, they decrease exponentially with increasing $|x|$.
This means that $x_{\rm L}$ and $x_{R}$ in Eqs.\,\eqref{wkb} and \eqref{ret} can be chosen almost arbitrarily.
In the following calculations we set $-x_{\rm L}\=\frac{1}{2}L\=+x_{\rm R}$.

\begin{figure}[t]
\includegraphics[width=\linewidth]{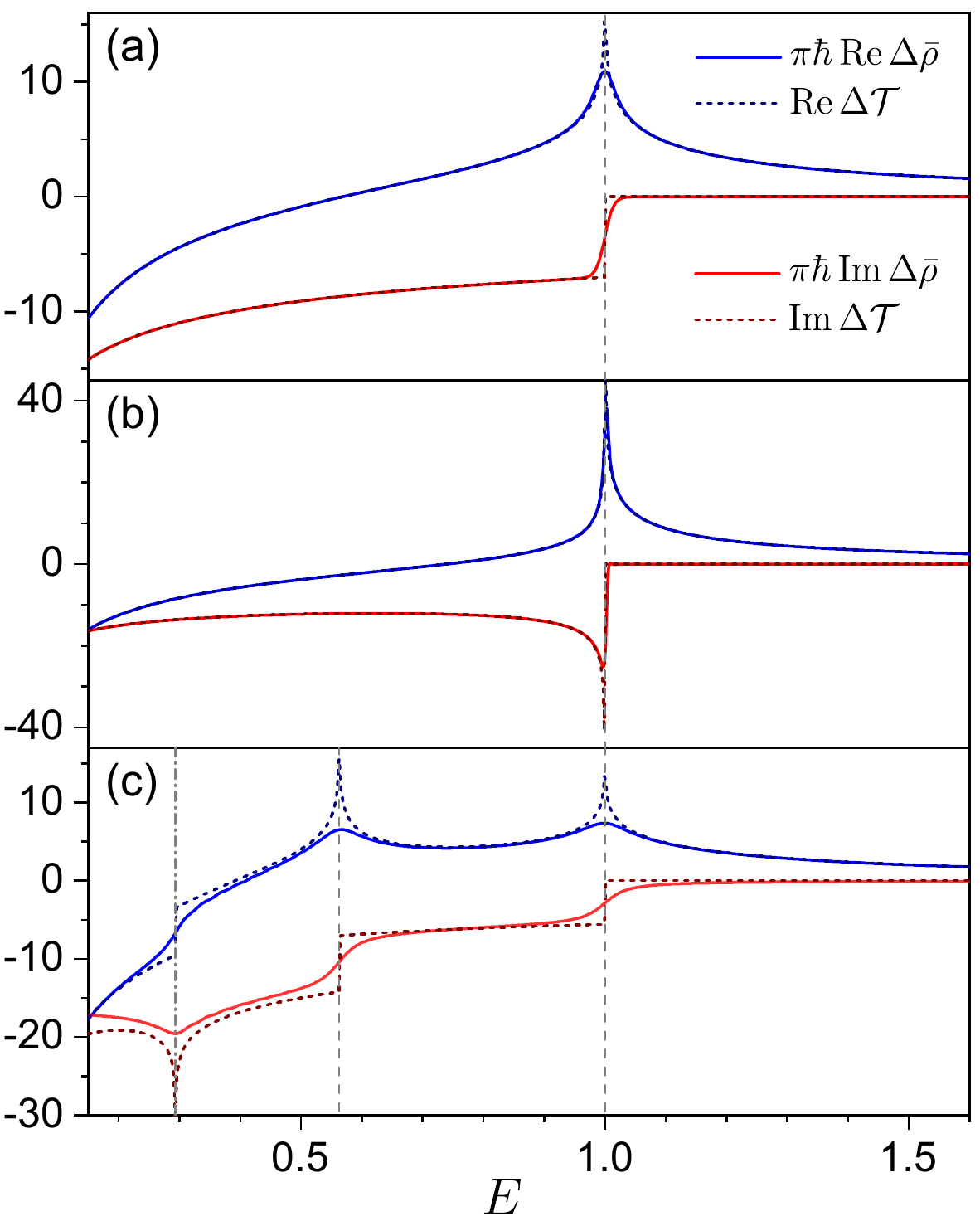}
\caption{
A test of formula \eqref{deti}.
Each panel shows the smoothed continuum level density (full curves) and the time shift (dashed curves) on the real energy axis for the potential in the respective panel of Fig.\,\ref{poti}.
The real and imaginary parts are drawn by blue and red curves, respectively.
Vertical lines mark energies of stationary points.
Parameters of the calculation are the same as in Fig.\,\ref{reso}, the smoothening parameter $\epsilon$ is $0.001$ in panels (a) and (b), and $0.050$ in (c).}
\label{time}
\end{figure}   

Figure~\ref{time} demonstrates the main result of this work.
It compares the real and imaginary parts of the smoothed continuum level density $\Delta{\overline\rho}(E)$ with the real and imaginary parts of the time shift $\Delta{\cal T}(E)$ from Eqs.\,\eqref{ret} and \eqref{imt}.
Note that the described method for the evaluation of $\Delta{\overline\rho}(E)$ fails close to $E\=0$, so the low-energy region is excluded.
Results in panels (a), (b) and (c) of Fig.\,\ref{time} refer to the potentials in the corresponding panels of Fig.\,\ref{poti}.

In accord with formula \eqref{deti}, we observe a satisfactory match of the $\Delta{\overline\rho}(E)$ and $\Delta{\cal T}(E)$ curves in Fig.\,\ref{time}. 
The agreement is expected to further improve with decreasing parameters $\hbar/\sqrt{m}$ (more resonances) and $\epsilon$ (less smoothening).
The match is very good in panels (a) and (b), and less good in panel (c), where the density curves are more smoothed as the corresponding potential gives a larger oscillatory component of $\Delta\rho(E)$.
In any case, the density and time curves in all panels of Fig.\,\ref{time} show the same qualitative features, particularly the step-, peak- and dip-like singularities at the energies associated with stationary points of the respective potentials.
These singularities reflect the fact that classical stationary points inside the interaction region induce anomalous changes of the complex time shifts.
As follows from the previous discussion, the singularities in ${\rm Re}\Delta{\cal T}(E)$ are connected with stationary points of $V(x)$, while those in ${\rm Im}\Delta{\cal T}(E)$ refer to stationary points in $-V(x)$.

Effects of the stationary points on the time shifts in Fig.\,\ref{time} can be classified as follows:

(i) A quadratic maximum of the potential at an energy $E_0$ leads to a logarithmic divergence $\propto-\ln|E\-E_0|$ of the time shift \cite{Str14+}.
This concerns the maximum of $V(x)$ in panel (a) of Fig.\,\ref{poti}, and the two maxima of $V(x)$ and one maximum of $-V(x)$ in panel (c).  
So in panel (a) of Fig.\,\ref{time} we observe one divergence in ${\rm Re}\Delta{\cal T}(E)$, while in panel (c) we get two divergences in ${\rm Re}\Delta{\cal T}(E)$ and one in $-{\rm Im}\Delta{\cal T}(E)$. 
Note that if the singularity appears in imaginary time, it is inverted in both time and energy directions since in the forbidden regions we do transformation $E\!\to\!-E$ and let time pass in the $-i$ direction.

(ii) A quadratic minimum of the potential at $E\=E_0$ produces a step-like dependence $\propto\Theta(E\-E_0)$ of the time shift (where $\Theta$ is a step function equal to 0 for negative arguments and to 1 otherwise) \cite{Str14+}.
This is the case of all structures in Fig.\,\ref{poti} mentioned in item (i), but in the inverse sense.
%---one minimum of $-V(x)$ in panel (a), and one minimum of $V(x)$ and two minima of $-V(x)$ in panel (c).
So in panel (a) of Fig.\,\ref{time} we have one step singularity of $-{\rm Im}\Delta{\cal T}(E)$, and in panel (c) two step singularities of $-{\rm Im}\Delta{\cal T}(E)$ and one of ${\rm Re}\Delta{\cal T}(E)$. 

(iii) A degenerate (higher than quadratic) extreme of the potential causes divergent time shifts in both minimum and maximum cases \cite{Str14+}.
This concerns the flat potential in Fig.\,\ref{poti}(b).
A quartic maximum of $V(x)$ at $E_0$ leads to a power-law divergence ${\rm Re}\Delta{\cal T}(E)\propto|E\-E_0|^{-1/4}$.
The corresponding quartic minimum of $-V(x)$  gives rise to the dependence $-{\rm Im}\Delta{\cal T}(E)\propto\Theta(E_0\-E)|E\-E_0|^{-1/4}$.

All the above singularities are reproduced in Fig.\,\ref{time} by the curves ${\rm Re}\Delta{\cal T}(E)$ and ${\rm Im}\Delta{\cal T}(E)$, and their correlates are seen in the associated  curves ${\rm Re}\Delta{\overline\rho}(E)$ and ${\rm Im}\Delta{\overline\rho}(E)$.
We know that singular energy dependencies of the level density in bound quantum systems are connected with the ESQPTs, which for systems with a single degree of freedom $f$ originate in non-analytic variations of classical periods of closed orbits \cite{Cej06,Cap08,Lar13,Str14+}. 
The present analysis therefore generalizes the concept of the $f\=1$ ESQPT from bound to scattering systems.
It shows that in the latter case dual ESQPT structures exist separately in both real and imaginary parts of the level density.
They are classified by stationary points of the normal and inverted potential applied in allowed and forbidden regions, respectively.
We point out that semiclassical calculations of tunneling resonances---already revealing an anomaly connected with the barrier maximum---were reported earlier in Refs.\,\cite{Con68+,Roy78}. 
Our work extends these results to general stationary points of any tunneling potential.

In summary, we have investigated the description of 1D scattering processes in terms of the continuum level density.
We have extended the existing definition of $\Delta\rho(E)$ to the complex domain and related its real and imaginary parts on the real energy axis to the real and imaginary time shifts associated with complex tunneling trajectories.
Fundamental relations \eqref{deriphas} and \eqref{deti} have been proposed.
As a confirmation of our surmise, we have clearly identified singularities of $\Delta{\overline\rho}(E)$ caused by classical stationary points for several test potentials $V(x)$.
In particular, the singularities in ${\rm Re}\Delta{\overline\rho}(E)$ reflect stationary points of $V(x)$ in the classically allowed regions, while the singularities in ${\rm Im}\Delta{\overline\rho}(E)$ reflect stationary points of $-V(x)$ in classically forbidden regions.
These dual structures constitute a remarkable generalization of ESQPTs to unbound systems with $f\=1$.
We anticipate that these results extend to $f\>1$ systems in which the ESQPTs affect higher derivatives of the level density \cite{Str14+,Str16}.
In general, our findings highlight the pertinence of instanton-like solutions of the tunneling problem and
contribute to the evidence that classical mechanics extended to the complex time domain provides valid semiclassical counterparts of purely quantum processes.

Let us stress that current nanoscience allows to realize customized resonant tunneling potentials via fabricating suitable nanostructures, see e.g. Refs.\,\cite{See01,Bha06,Suz10,Bri13,Gol15,Tao19}.
So the above ESQPT effects, translated through Eqs.\,\eqref{tra} and \eqref{deriphas} to variations of the tunneling amplitude, are of direct practical importance and open to experimental tests.

We thank N. Moiseyev for inspiring lectures on the complex scaling method during Prague AQM1 summer school in 2019.
The work was supported by the Czech Science Foundation (grant nos.\,20-09998S and 20-21179S) and by Charles University (UNCE/SCI/013).

\end{document}